\documentclass[amssymb, amsmath, aps, prx, reprint, showkeys, showpacs]{revtex4-1}

\usepackage{graphicx}

\usepackage[colorlinks=true,linkcolor=blue]{hyperref}
\expandafter\ifx\csname package@font\endcsname\relax\else
 \expandafter\expandafter
 \expandafter\usepackage
 \expandafter\expandafter
 \expandafter{\csname package@font\endcsname}
\fi
\usepackage{color}

\begin{document}


\title{Predicting kinetics of polymorphic transformations from structure mapping and coordination analysis}


\author{Vladan Stevanovi\'c}
\email{vstevano@mines.edu}
\affiliation{Colorado School of Mines, Golden, Colorado 80401, USA}%
\affiliation{National Renewable Energy Laboratory, Golden, Colorado 80401, USA}%

\author{Ryan Trottier}
\affiliation{University of Colorado Boulder, Colorado 80309, USA}%

\author{F\'elix Therrien}
\affiliation{Colorado School of Mines, Golden, Colorado 80401, USA}%

\author{Charles Musgrave}
\affiliation{University of Colorado Boulder, Colorado 80309, USA}%

\author{Aaron Holder}
\affiliation{National Renewable Energy Laboratory, Golden, Colorado 80401, USA}%
\affiliation{University of Colorado Boulder, Colorado 80309, USA}%

\author{Peter Graf}
\affiliation{National Renewable Energy Laboratory, Golden, Colorado 80401, USA}%

\date{\today}

\begin{abstract}
To extend rational materials design and discovery into the space of metastable polymorphs, rapid and reliable assessment of their lifetimes is essential. 
Motivated by the early work of Buerger (1951), here we investigate the routes to predict kinetics of polymorphic transformations using solely crystallographic 
arguments. As part of this investigation we developed a general algorithm to map crystal structures onto each other and construct transformation 
pathways between them. The developed algorithm reproduces reliably known transformation pathways and reveals the critical role that the 
dissociation of chemical bonds along the pathway plays in choosing the best (low-energy barrier) mapping. By utilizing our structure-mapping algorithm we were able to 
quantitatively demonstrate the intuitive expectation that the minimal dissociation of chemical bonds along the pathway, or in ionic systems, the condition of coordination 
of atoms along the pathway not decreasing below the coordination in the end compounds, represents the requirement for fast transformation kinetics. We also demonstrate 
the effectiveness of the structure-mapping algorithm in combination with the coordination analysis in studying transformations between polymorphs for which
a detailed atomic-level picture is presently elusive.
\end{abstract}
%
%
\maketitle
%
\section{Introduction}
Strong dependence of physical properties on crystal structure offers compelling opportunities for discovering new
functional materials among metastable polymorphs. The search, however, critically requires:  
({\it i}) accurate and efficient structure and property prediction algorithms to screen the potential energy surface (PES) for polymorphs with 
attractive properties, and ({\it ii}) rapid and reliable assessment of the kinetics of transformations to lower energy states to help identify 
long-lived polymorphs among the multitude of possible low-energy structures. 
The former problem has attracted the attention of the scientific community 
due to what was initially thought to be insurmountable challenges associated with predicting the structure of periodic systems. As a result, 
a range of structure prediction methods have been developed \cite{catlow_NMAT:2008, oganov2010modern, evrenk2014prediction} and 
successfully combined with property predictions \cite{hautier_CM:2011, lyakov_PRB:2011, botti_PRB:2012, li_JCP:2014, peng_PRX:2015, 
anubhav_NRM:2016, pickard_APLM:2016, dkvashnin_JPCL:2017}. In the context of metastability and polymorphism,
identification of bonafide metastable structures among tens if not hundreds of low-energy 
structures that typically result from structure predictions\cite{zwijnenburg_PRL:2010,zwijnenburg_PRB:2011} still remains a challenge\cite{Sune1600225}. 
The recent discovery of a correlation between volumes of configuration space occupied by different PES local minima (their basins of attraction) 
and experimentally realized metastable structures \cite{sandip_PRL:2014,vladan_PRL:2016,jones_arxiv:2017} offers some help in narrowing down the list 
of candidate metastable structures. However, when targeting metastable forms of matter, knowledge of the kinetics of transformations to lower 
energy states remains invaluable.\par
 
Efforts in developing approaches to predict the minimal energy transformation pathways and associated structures and energies of transitions states, 
to assess the kinetics of polymorphic transformations have also been (and still are) under development. The most common is probably the generalized 
solid-state nudged elastic band (ssNEB) formalism \cite{Caspersen10052005, sheppard_JCP:2012, Qian20132111}, a technique for locating saddle-points 
on the potential energy surface to determine the lowest-energy transformation pathways (trajectories of atoms) between different crystal structures, and to 
compute the corresponding energy barriers. The ssNEB is a generalization of the original NEB method \cite{henkelman_JCP_1:2000, henkelman_JCP_2:2000} 
developed for non-periodic systems. The main challenge in applying the ssNEB method is choosing the best initial pathway 
that will then be modified by the algorithm to the closest minimum energy path.\par

Despite the impact that these developments have had on determining mechanisms and predicting kinetics, they are not suitable for the screening of 
large number of structures for metastable polymorphs. A method that could examine large sets of polymorphs, even if offering only a qualitative 
assessment of the kinetics of polymorphic transformations, would greatly benefit metastable polymorph discovery.
One such qualitative assessment can be found in the work of Buerger \cite{buerger1951}, who classified polymorphic transformations 
as {\it slow} or {\it rapid} using mainly crystallographic arguments. In short,
the existence of a diffusionless (displacive or dilatational) transformation between two structures is, according to Buerger,
a signature of fast kinetics. One transformation that disobeys this classification is diamond to graphite 
for which a relatively simple, diffusionless pathway can be constructed \cite{parinello_NM:2011,henkelman_D2G_JCP:2012}, 
but diamond is a known long-lived allotrope of carbon at ambient conditions. To account for this, Buerger devised an 
additional class of transformations he called bond type, covalent to metallic in this case, which he argued should be slow
irrespective of geometry.\par

\begin{figure}[t!]
\includegraphics[width=\linewidth]{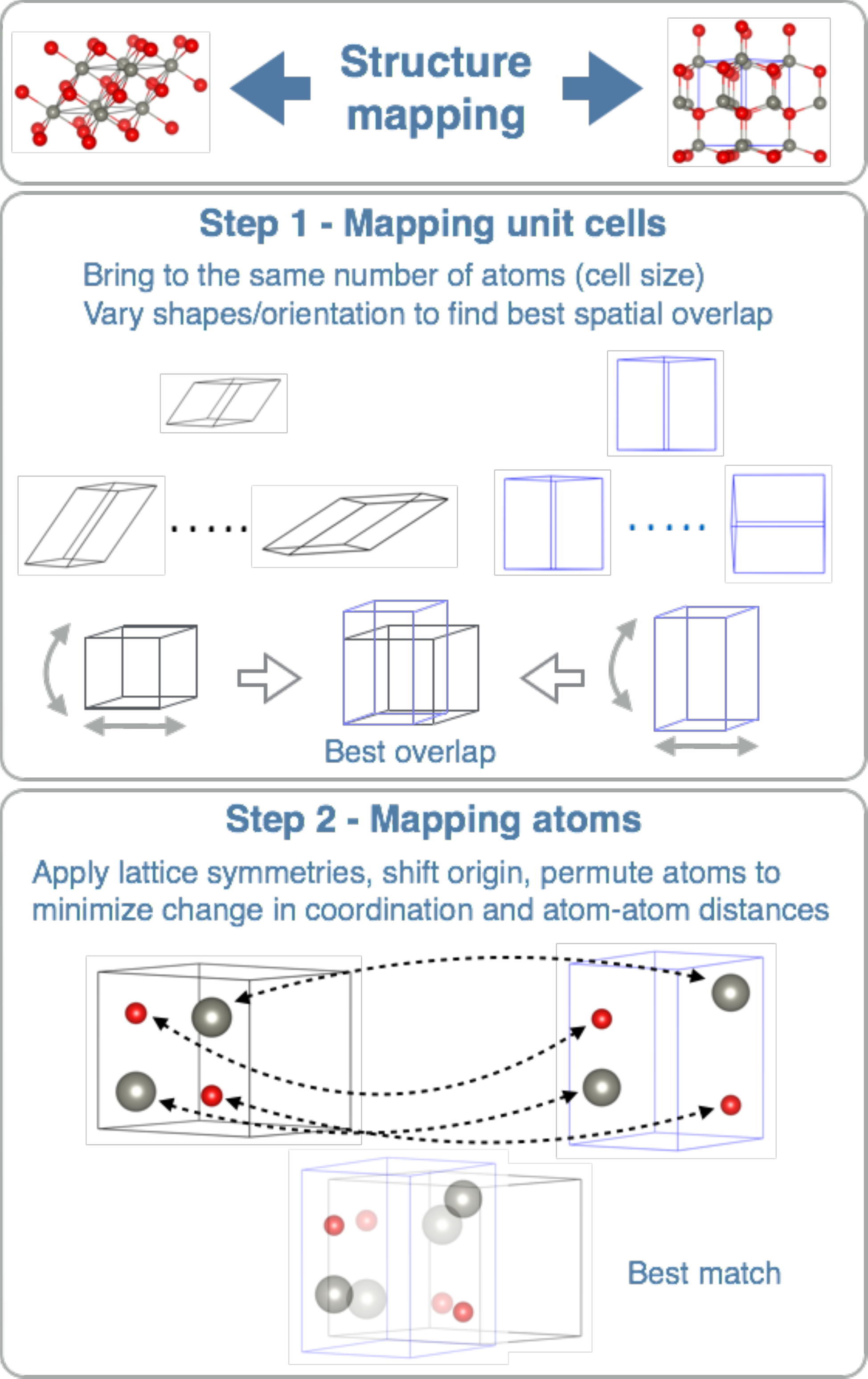}
\caption{\label{fig1} 
Schematics of the two step, structure mapping algorithm developed in this work (see text for details).
}
\end{figure}

Motivated by the Buerger's work, we investigate in this paper the routes to generalize his classification and convert it from a collection of heuristics into a 
computational approach with the goal of accelerating the assessment of the kinetics of polymorphic transformations. The key component of our work is 
an algorithm to map crystal structures onto each other and construct diffusionless transformations between them (see Fig.~\ref{fig1}). The mapping 
algorithm relies on two basic principles: (a) minimization of the total Euclidian distance atoms need to travel between the end structures, and (b) minimization 
of the change in coordination of atoms in the first coordination shell (number of atoms in the first shell) along the map. In other words, the goal is to find an 
optimal mapping that is diffusionless in nature and which minimizes the dissociation of chemical bonds. \par

We will show that the developed algorithm successfully reproduces known transformation pathways in well studied systems including $fcc$ to $bcc$, diamond to graphite, 
CsCl-type to rocksalt, etc. It reveals the critical role of minimizing the change in coordination in searching for the best mapping. Beyond simple and well studied systems, 
we applied our mapping algorithm to polymorphic transformations between various SnO$_2$ high-pressure phases. Our results show that if only a qualitative assessment 
of the kinetics of polymorphic transformations is needed, the condition of minimal dissociation of chemical bonds along the pathway, can be used as a signature of fast kinetics. 
Note that, in partially ionic systems, the condition of minimal dissociation, can be formulated as the coordination of atoms in the first shell not decreasing below the coordination 
in the end structures. Lastly, if a more accurate estimate of the transition pathway and the associated energy profile is necessary, our mapping algorithm is demonstrated to provide a 
good starting point for subsequent solid-state nudged elastic band (ssNEB) calculations. \par

Similarly to other methods \cite{Capillas_JPCM:2007, Caspersen10052005, sheppard_JCP:2012, Qian20132111}, 
in our work we also draw conclusions about the kinetics of polymorphic transformations from the assumed collective (concerted) motion of atoms.
While phase transitions in solids occur mainly via nucleation and growth, we adopt the view that the features of collective transformations 
can provide useful insights into the overall kinetics and can serve as the starting point for more realistic modeling of nucleation processes. \par
%
%
%
\section{Structure mapping algorithm}

The developed structure mapping algorithm consists of two steps as illustrated in Fig.~\ref{fig1}.
In the first step the algorithm searches for the most compatible representations of the two structures, that is 
the optimal mapping of their unit cells.
This is done in the following way. The two primitive unit cells are brought to the same -- least common multiple -- number 
of atoms by constructing all symmetry inequivalent supercells. The enumeration of the symmetry inequivalent supercells given the 
number of atoms is done using Hart-Forcade theory \cite{hart_PRB:2008}. The most compatible representations are 
then defined by the pair of supercells that minimizes the strain between them, or, in other words, the two supercells with the largest 
spatial (volumetric) overlap. This condition can be formulated as searching for the pair of unit cells that minimizes the 
metric defined as the weighted sum of the absolute differences in cell parameters $(a, b, c, \alpha, \beta, \gamma)$ and 
the total surface areas (S) of the two cells:
\begin{equation}\label{eq:celldist}
d(cell_1,cell_2)=\sum_{\substack{q=a, b, c, \\ \alpha, \beta, \gamma, S}} \, C_q \, | q_1 - q_2 |,
\end{equation}
where $C_q$ represent positive weights of quantities $q$ that are introduced solely to make the numerical values the same order of magnitude. 
The search is accomplished by transforming the cells to the corresponding (unique) reduced cell according to the formulation of Niggli-Santora-Gruber 
\cite{Niggli_ExptPhys:1928,Santora_Acta:1970,Gruber_Acta:1976}, which allows implicitly exploring all permutations of the unit vectors and all isometric 
transformations of the two cells (rigid rotations and reflections) to bring them to the positions of largest spatial overlap, as shown in Fig.~\ref{fig1}.\par

In the second step, the atoms are placed back inside the two cells. The optimal atom-to-atom mapping is then 
searched for by performing the following operations on the two sets of atomic positions: ({\it i}) all symmetry operations 
of the parent Bravais lattices, ({\it ii}) translations of the origin of coordinate frame to each 
atom site, and ({\it iii}) permutations of indices of chemically identical atoms. Every choice of symmetry operation, 
position of the origin, and the permutation of atom indices defines a single mapping between the two end structures and a 
pathway in configuration space that connects the atoms with same indices and continuously deforms the unit cell.\par

Out of many possible atom-to-atom mappings our algorithm selects as the optimal mapping the solution that yields minimal dissociation of chemical bonds, i.e., 
the coordination of atoms along the pathway does not decrease below the coordinations in the end structures. If such a mapping cannot be found, then the one that 
minimizes the sum of Euclidian distances between the corresponding atoms in the two structures is chosen. Finally, if more than one solution is found with the appropriate 
change in coordination then the sum of distances between the atoms is used to rank them and narrow down the choice. To achieve this, the described 
operations are performed using Hart-Forcade theory \cite{hart_PRB:2008} that allows enumeration of the symmetry inequivalent atom sites. The Hungarian Algorithm of 
Kuhn and Munkres \cite{kuhn_hungarian:1955} is used to find the optimal permutation of atom indices (job scheduling problem) in polynomial time.\par

Ideally, the globally best atom-to-atom mapping would always be found in the unit cell pairing with highest overlap, but this is not the case in general.
Therefore, in the first step, we form a list of the top $N$ unit cell pairing (e.g. $N$=10) and perform the second step on all of them.
In practice this has been found to yield the desired globally optimal atom-to-atom mapping.\par

A similar approach was developed previously by Sadeghi and Goedecker \cite{sadeghi_JCP:2013} for the purpose of measuring configuration space distances between non-periodic 
systems. Also, Lonie and Zurek \cite{lonie_CPC:2012} developed a search algorithm designed to identify identical (duplicate) periodic structures which involves mapping of the unit cells. 
Another class of recently developed approaches utilize the descriptor/feature based fingerprinting to quantify similarity of different periodic structures by comparing selected 
set of features (not atom-by-atom). These include the work of Yang et al. \cite{yang_PRB:2012} and Zhu et al. \cite{zhu_JCP:2016}. Our mapping of the unit cells 
can be viewed as a generalization of the ideas of Lonie and Zurek \cite{lonie_CPC:2012} to the case where input structures are presumed to be different and where the goal 
is to discover the optimal alignment of the two structures. Concerning the atom-to-atom mapping, we extended the algorithm of Sadeghi and Goedecker \cite{sadeghi_JCP:2013} 
to periodic systems. Another important distinguishing feature of our approach is the new objective function, which includes the physical principle of the minimal dissociation of 
chemical bonds. \par

We tested our structure mapping algorithm on a number of well-studied polymorphic transformations in elemental and simple binary systems, 
some of which are listed in Table~\ref{tab:sgs}. The space groups of the lowest symmetry structures that occur along the pathway are also shown. 
The transformation pathways from  Table~\ref{tab:sgs} reproduce 
well the maximal symmetry transition paths for the reconstructive phase transitions derived by Capillas et al. \cite{Capillas_JPCM:2007}. 
Further, our algorithm often finds the pathway that preserves the largest common subgroup of the two structures. 
This is a consequence of the supercell sizes chosen to accommodate the least common multiple number of atoms. 
If this condition is relaxed and larger supercells with compatible numbers of atoms are considered, our algorithm would also find 
lower symmetry pathways, some of which have been discussed previously for the transformations from 
Table~\ref{tab:sgs}. The details of the transformations from Table~\ref{tab:sgs} are provided in the supplementary materials. 
%
\begin{table}[t!]
\caption{\label{tab:sgs}
List of studied polymorphic transformations in elemental and simple binary systems. Space groups of the
initial and final structures are given together with lowest symmetry intermediate structures.}
\begin{ruledtabular}
\begin{tabular}{rcccl}
FCC (Fm$\bar{3}$m)         & $\rightarrow$ & I4/mmm & $\rightarrow$ & BCC (Im$\bar{3}$m) \\
BCC (Im$\bar{3}$m)          & $\rightarrow$ & Cmcm   & $\rightarrow$ & HCP (P6$_3$/mmc) \\
diamond (Fd$\bar{3}$m)   & $\rightarrow$ & C2/m     & $\rightarrow$ & graphite (P6$_3$/mmc) \\
CsCl-type (Pm$\bar{3}$m) & $\rightarrow$ & R$\bar{3}$m & $\rightarrow$ & rocksalt (Fm$\bar{3}$m) \\
rocksalt (Fm$\bar{3}$m)   & $\rightarrow$ & Cmc2$_1$ & $\rightarrow$ & wurtzite (P6$_3$mc) \\
zincblende (F$\bar{4}$3m) & $\rightarrow$ & Imm2     & $\rightarrow$ & rocksalt (Fm$\bar{3}$m) \\
\end{tabular}
\end{ruledtabular}
\end{table}
%
\section{Change in the coordination of atoms and its relevance to the magnitude of kinetic barriers}

Another important finding that emerged during the development of our algorithm is that the two conditions, minimal distance between 
the corresponding atoms in the two structures and minimal change in coordination, do not always coincide. In this section 
we analyze relatively simple transformations for which the coordination of atoms, {\it i.e}, the number of atoms in the first 
coordination shell, is relatively straightforward to define. A more through discussion on the definition of the first shell coordination is presented 
in Section~\ref{sno2}.\par

The test cases from Table~\ref{tab:sgs} revealed the critical significance of the change in coordination. For example, minimizing the distance alone 
does not necessarily result in the best pathway in the case of $bcc$ to $hcp$ transformation as there are multiple possible mappings
with nearly degenerate distance values, but different coordination of atoms along the pathway. The transformations between the ground state and high 
temperature tin-sulfide polymorphs, Pnma and Cmcm, also exhibit similar features (see supplementary materials). The condition itself has been referred to previously 
in case-by-case studies (e.g. Refs.~\cite{catti_PRL:2001, lambrecht_PRB:2003}) and has a relatively 
simple physical interpretation. It reflects the requirement of minimizing the number of broken bonds in going from one end 
structure to the other, which is expected to be correlated with the magnitude of the kinetic barriers. In ionic and covalent systems, 
because of the directionality of chemical bonds, this condition can be implemented just by requiring that the first-shell coordination 
of atoms along the pathway does not decrease below the coordination in the end structures.\par
%
%
\begin{figure}[t!]
\includegraphics[width=\linewidth]{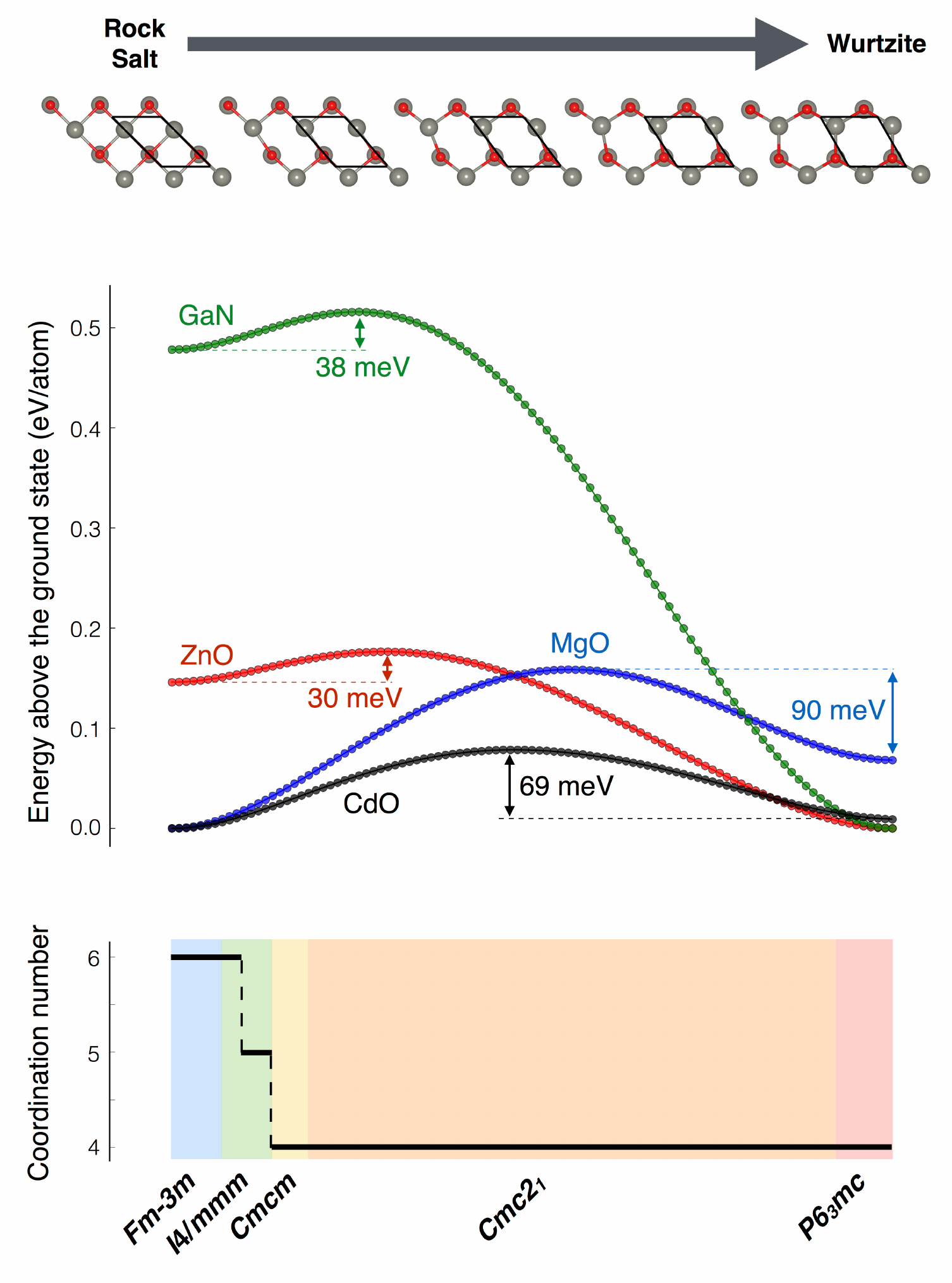}
\caption{\label{fig2} 
Five snapshots along the rocksalt (Fm$\bar{3}$m) $\rightarrow$ wurtzite (P6$_3$mc) transformation
are shown (upper) together with the energy profiles (middle) and the change in coordination
along the pathway (lower). The invisible $x$-axis represents normalized reaction coordinate.
}
\end{figure}

We illustrate the correlation between the change in the first shell coordination and the potential energy profile by analyzing the 
rocksalt (Fm$\bar{3}$m) $\rightarrow$ wurtzite (P6$_3$mc) transformation in binary ionic systems GaN, ZnO, MgO, and CdO. In the upper part of 
Fig.~\ref{fig2} five snapshots along the pathway are shown viewed from the wurtzite $c$-axis. In the middle part, potential energy 
profiles along the pathway are shown. They are calculated using density functional theory (DFT).  The energy axis is relative to each 
ground state structure, wurtzite for GaN and ZnO and rocksalt for MgO and CdO. For every chemical composition the end structures are 
fully relaxed (volume, cell shape and atomic positions), while the total energies of the snapshots along the pathway are computed without any relaxations. 
For each system the pathways are discretized into 100 equally spaced successive snapshots shown in Fig.~\ref{fig2}. We employ 
a relatively standard DFT-GGA numerical setup described in Section~\ref{methods}. \par

As evident from Fig.~\ref{fig2} for all four systems calculated energy profiles exhibit relatively small barriers for transformation from the 
higher energy into the corresponding ground state state structure, which corresponds well to the known, fast kinetics 
of this transformation \cite{decremps_APL:2002, lambrecht_PRL:2001}. The highest energy points along the pathways are all below 100 meV/atom, 
which is usually considered low activation energy \cite{trinkle_PRL:2003}. Because the calculated energy profiles can be thought of as providing the 
upper bounds for the true activation energies, this result shows that irrespective of the chemistry, this particular transformation is 
associated with relatively low energy barriers. In addition, the change in coordination of the atoms from the 6-fold coordinated rocksalt phase to 4-fold 
coordinated is evidently monotonic, i.e. only the absolutely necessary number of bonds dissociate along the pathway, which is consistent with the 
above discussion. Similar results are obtained for other transformations from Table~\ref{tab:sgs} including: BCC $\rightarrow$ HCP in metallic 
titanium for which we find a nearly barrierless energy profile in line with previous findings \cite{trinkle_PRL:2003}; CsCl-type $\rightarrow$ rocksalt in CsCl,
with the monotonic change in coordinations and the upper bound for the activation energy of 13 meV/atom. 
Moreover, for the zincblende $\rightarrow$ rocksalt transformation in SiC our algorithm confirms previously discussed
energetic preference of the orthorhombic (Imm2) pathway relative to the R3m mechanism because of the monotonic change in the coordination of 
atoms from four to six\cite{catti_PRL:2001, lambrecht_PRB:2003}. The calculated energy profiles peak at 112 and 250 meV for the pathways passing 
through the Imm2 and R3m, respectively, in agreement with previous studies.\par
%
\section{Diamond $\rightarrow$ graphite transformation in elemental carbon}
\begin{figure}[t!]
\includegraphics[width=\linewidth]{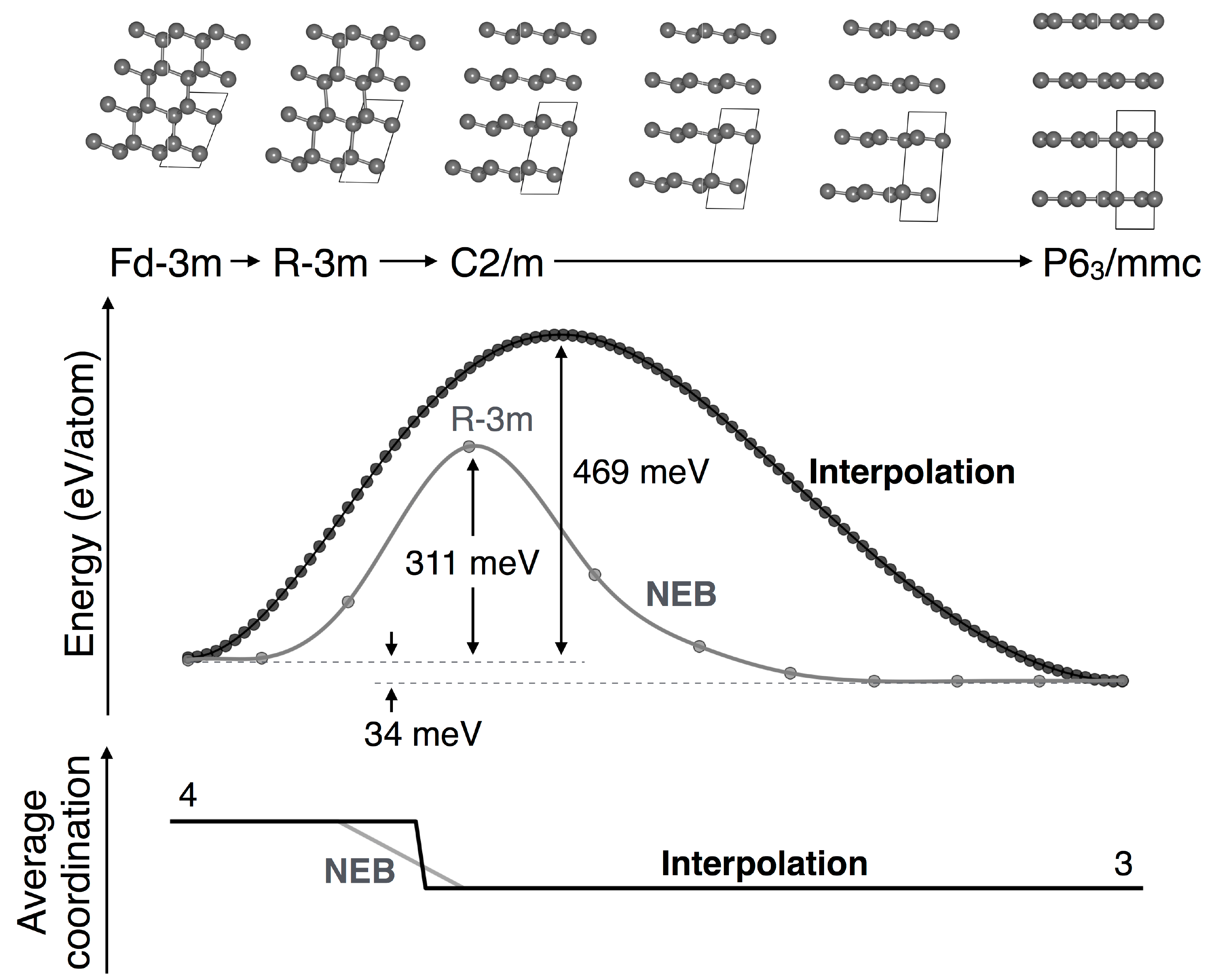}
\caption{\label{fig3} 
Snapshots along the diamond to graphite transformation are shown together with the space group symmetry, calculated energy profiles and
the evolution of the coordination number. Two energy profiles calculated using the discretized pathway produced by our algorithm (black) and
the ssNEB result (grey) are shown as well as the corresponding coordination numbers colored in the same way.}
\end{figure}

%
Our results also confirm the exception to the rule, the cubic diamond $\rightarrow$ hexagonal graphite transformation for which our algorithm finds
the same transformation pathway as the one discussed in Refs.~\cite{parinello_NM:2011,henkelman_D2G_JCP:2012} that is accompanied by a monotonic change in coordination 
of atoms (see Fig.~\ref{fig3}) and yet, the maximal energy point along the pathway is $\sim$ 470 meV/atom above the diamond structure. Similar to other systems studied herein, 
the calculations were performed by discretizing the pathway constructed by our algorithm (interpolation). Fig.~\ref{fig3} also illustrates the energy profile in grey, which resulted 
from the ssNEB calculation that used our mapping algorithm pathway as the initial pathway. The energy profiles are calculated using the optB86 exchange-correlation functional that 
includes contributions coming from Van der Waalls interactions  \cite{klimes_JPCC:2010} as implemented in the VASP software package \cite{kresse_CMS:1996}. 
As expected, the calculated ssNEB energy barrier is lower, but the qualitative picture does not change. The transformation is slow despite 
the monotonic change in coordination reproduced by both sets of calculations.\par
%

So, why is elemental carbon different?  As correctly observed by Buerger, the diamond to graphite transformation requires a change in chemical bonding. In the diamond structure 
the four carbon $sp^3$ hybrid atomic orbitals form four strong $\sigma$ bonds per carbon atom. These are replaced in the graphite structure by three 
$sp^2$ hybrids, which form three $\sigma$ bonds per carbon, and a $p_z$ orbital, which forms one weaker, resonating $\pi$ bond per carbon. Relatively weak Van der Waals 
interactions between the layers further stabilize the graphite structure. If one counts the number of chemical bonds per atom rather than the geometric coordination, 
then both diamond and graphite have four bonds per carbon atom and the transformation from diamond to graphite would imply dissociation and formation of one bond per C. 
Consistent with the previous discussion, the intermediate decrease in the number of chemical bonds to three from four in the end structures would lead to a high barrier.\par

In ionic systems on the other hand, significant contribution to the energy differences between different atomic configurations 
comes form purely electrostatic interactions \cite{vladan_PRL:2010, vladan_JACS:2011}. Hence, the increase in energy along the pathway is influenced by the changes
in the charge distribution and to a lesser extent is due to vanishing overlaps of atomic overlaps. The evidence of the remaining charge transfer along the pathways 
are the non-vanishing band-gaps for all ionic systems studied here. Therefore, the argument here is that the geometric coordination of atoms along the pathway 
is more appropriate when trying to understand the kinetics of polymorphic transformations in ionic systems, while in covalent systems it needs to be replaced by a 
chemical bonding analysis. In both cases however, the condition of minimal dissociation of chemical bonds serves as a signature of rapid transformations. \par
%
\section{SnO$_2$ polymorphs}\label{sno2}
%
To further validate the previous discussion we extend our study to SnO$_2$, a partially ionic system for which a number of polymorphs have been 
realized under pressure. With increasing pressure the structures appear in the following sequence: P4$_2$/mnm $\rightarrow$ Pnnm $\rightarrow$ Pbcn 
$\rightarrow$ Pa3 $\rightarrow$ Pbca $\rightarrow$ Fm$\bar{3}$m \cite{shieh_PRB:2006, das_PMS:2014}. Upon releasing the pressure, however, all phases 
either relax back to the ground state rutile (P4$_2$/mnm) structure following the same sequence or to a phase mixture between rutile and the Pbcn structure
($\alpha$-PbO$_2$ structure type) \cite{das_PMS:2014, liu_Science:1978}. So, the only phase that survives at ambient conditions is 
P4$_2$/mnm, occasionally in combination with small amounts of Pbcn. A previous study \cite{vladan_PRL:2016} has shown that these two SnO$_2$ structures,
P4$_2$/mnm and Pbcn, have the ``largest'' local minima, that is, they occupy larger regions of configuration space than any other. Here, we extend this result by 
investigating transition pathways between different SnO$_2$ polymorphs.\par

\begin{figure}[t!]
\includegraphics[width=\linewidth]{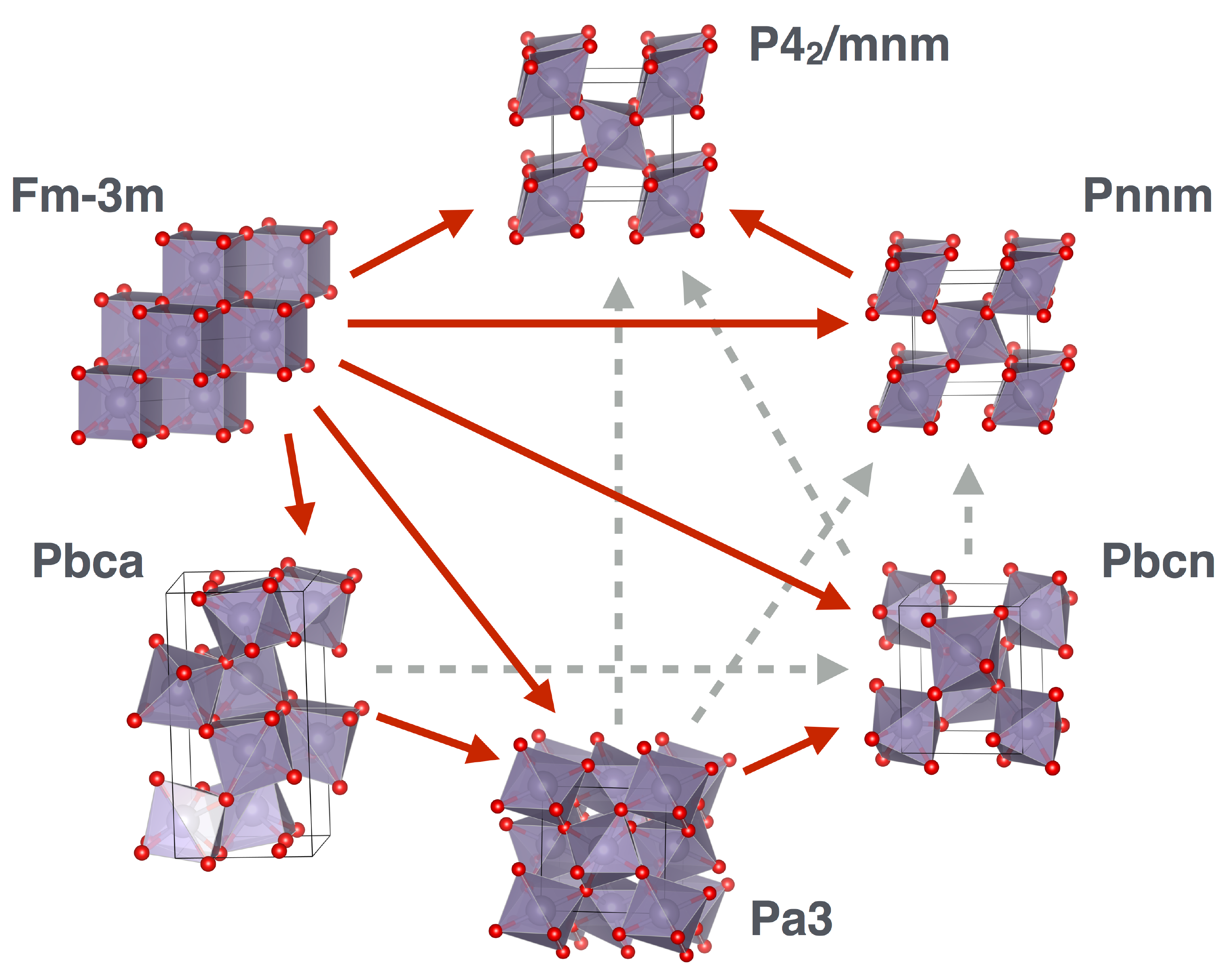}
\caption{\label{fig4} 
Map of six SnO$_2$ polymorphs. Thick arrows indicate {\it fast} polymorphic transformations as predicted by
our structure mapping and coordination analysis. Arrows go in the direction of lowering energy. 
Dashed arrows indicate transformations for which the change in coordination depends strongly on the 
distance cutoff used (see text). The missing arrows correspond to transformations predicted to be {\it slow}.
}
\end{figure}

In Fig.~\ref{fig4}, a chart illustrating the crystal structures of all six SnO$_2$ polymorphs is shown with thick arrows connecting structures for which 
our structure mapping and coordination analysis suggest fast polymorphic transformations. The arrows point in the direction of lowering 
total energy. Interestingly, the highest pressure Fm$\bar{3}$m phase is connected to all other phases by a fast polymorphic transformation. The DFT 
calculated energy profiles all exhibit energy barriers lower than 30 meV/atom. \footnote{Complete information about all SnO$_2$ polymorphs, their 
relative energies as well as the energy profiles for the transformations between them are provided in Supplementary Information.} Hence, upon 
releasing the pressure the Fm$\bar{3}$m structure can, depending on the actual barriers and other factors such as how fast the pressure is 
released or defects in the material, transform relatively quickly to any of the other SnO$_2$ polymorph structures upon releasing pressure. 
This semi-quantitative assessment is consistent with the observations from the high-pressure experiments.\cite{das_PMS:2014, liu_Science:1978}\par
%
%
\begin{figure*}[t!]
\includegraphics[width=\linewidth]{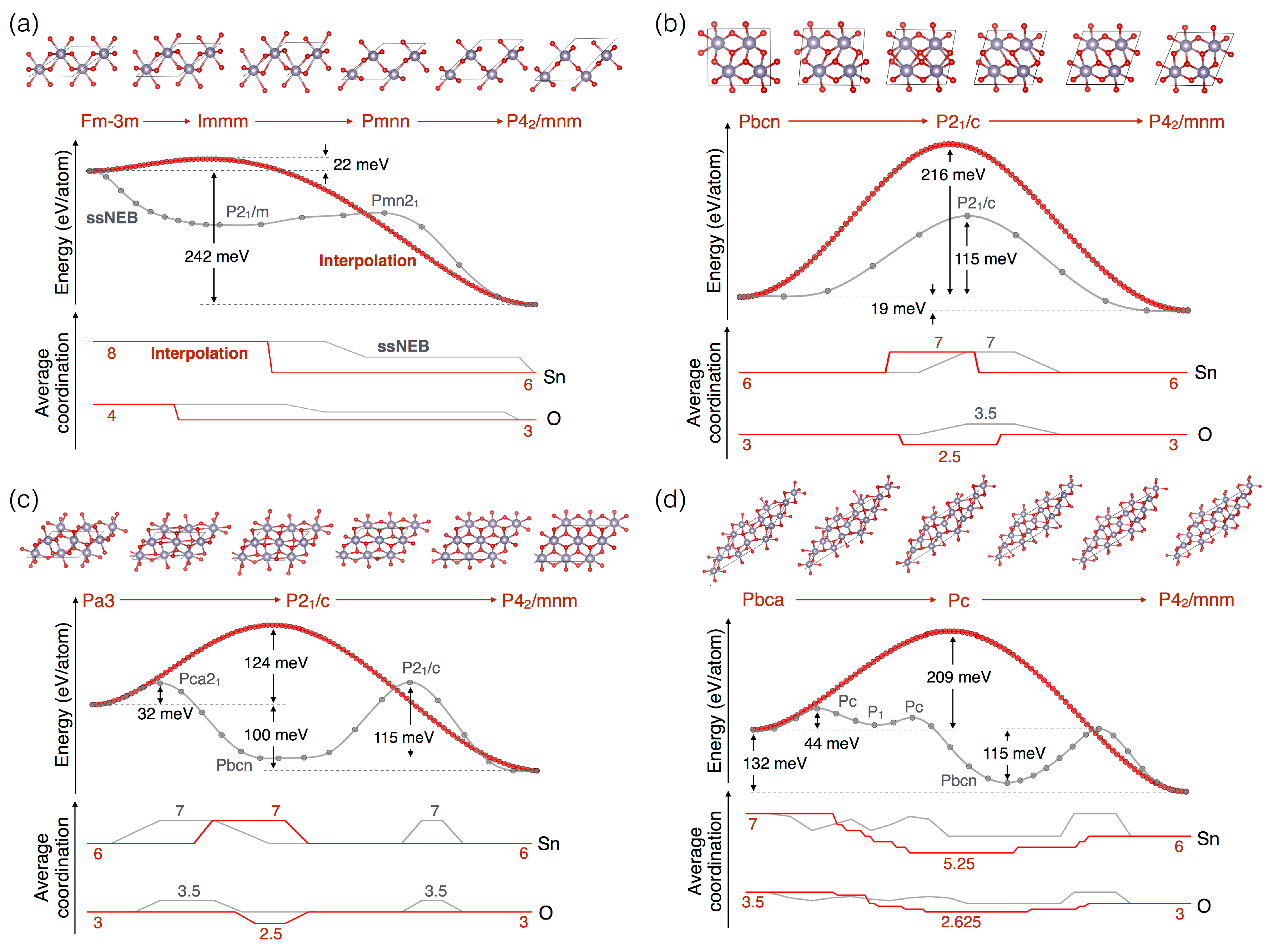}
\caption{\label{fig5} 
Calculated energy profiles are shown (in red) for the four selected polymorphic transformations from Fig.~\ref{fig4} 
together with the space group symmetry, five crystal structure snapshots along the pathway and the average coordination of Sn and O. The 
corresponding ssNEB results are shown in grey. The $x$-axis represents the normalized reaction coordinate.}
\end{figure*}
%
We also find another classe of transformations between the SnO$_2$ polymorphs, 
those marked in Fig.~\ref{fig4} by dashed arrows. For these transformations our classification based on 
the coordination analysis is very sensitive to how the coordination of atoms is computed. Namely, along the pathways the symmetry of the end structures 
is usually broken, and the distances between the atoms and their first neighbors are not all the same. As a result, the calculated number of neighbors 
will in some cases depend on the cutoff radii and/or tolerances on these distances.  For binary ionic systems we define the first coordination shell as 
consisting of the atoms of the type other than the central atom that are all separated from the central atom within a certain tolerance factor. 
In other words, the tolerance factor represents the thickness of a spherical shell around the central atom. The first shell is then defined as the number of atoms of the other type 
that are found inside this shell. The counting starts from the closest atom and stops if either the tolerance or the atom of the same type as the central one is reached, whichever comes first. 
We check the stability of all our results by varying the tolerance factor from 0.1 to 0.5 {\AA}. \par

The dashed arrows in Fig.~\ref{fig4} represent the transformation pathways that would either be classified as slow for the smallest tolerance 
factor (0.1 {\AA}) and would change to fast upon increasing the tolerance, or for which the results would be ambiguous, {\it i.e.,} the trends in the coordination
of different types of atoms would be different (the opposite). The former physically means that along the pathway 
some chemical bonds are strained more than others, but the atoms remain relatively close to each other. The latter means that the atoms of the same type 
as the central atom approached closer to the central atom and intermixed with its first coordination shell and in that way decreased the evaluated coordination number. 
Based on the subsequent ssNEB calculations we argue that it is more appropriate to classify these types of transformations as likely rapid for the following reason. 
Namely, our procedure is based solely on geometry and does not
allow atoms to relax to more stable configurations as, for example, the ssNEB method would. Therefore, because of the energy minimization with respect to atomic positions 
within the ssNEB the atoms will have the opportunity to re-bond during the atomic relaxations, which would then lower the barrier and imply fast transformation
based on the coordination analysis. Based on this discussion, all dashed-line transformations from Fig.~\ref{fig4} should be classified as rapid.\par

Finally, the coordination analysis along the Pbca $\rightarrow$ P4$_2$/mnm and Pbca $\rightarrow$ Pnnm pathways clearly shows the dissociation of chemical bonds
and consequently, we do not connect these structures with arrows. Unlike those marked by the dashed arrows, the change in coordination along  Pbca $\rightarrow$ P4$_2$/mnm 
and Pbca $\rightarrow$ Pnnm does not depend on the details of how is the first shell coordination of atoms evaluated.\par

To better illustrate previous discussion we show in Fig.~\ref{fig5} calculated energy profiles and the average coordination of atoms along the pathways together with the 
ssNEB results for a selected set of transformations from Fig.~\ref{fig4}. We consider Fm$\bar{3}$m $\rightarrow$ P4$_2$/mnm (thick arrow), Pbcn $\rightarrow$ P4$_2$/mnm 
(dashed arrow), Pa3 $\rightarrow$ P4$_2$/mnm (dashed arrow), and Pbca $\rightarrow$ P4$_2$/mnm (no arrow). The first transformation, which would be classified as 
fast according to our coordination analysis, clearly indicates the existence of a low energy barrier. The highest energy point along the pathway is only 22 meV/atom above 
the high-pressure Fm$\bar{3}$m phase. The transformation is nearly barrierless in the ssNEB showing a very similar change in the coordination of atoms. The second and 
third transformations that are denoted by dashed arrows in Fig.~\ref{fig4} both fall in the category of undetermined based on the initial coordination analysis. Namely, the coordination 
numbers for Sn and O follow opposite trends. While the coordination of Sn grows along the pathway from six to seven the average number of first shell Sn atoms 
surrounding oxygen drops from 3 to 2.5. The reason for this is already mentioned intermixing of O atoms within the first shell of other O atoms,
which decreases their first shell coordination number (as defined here) and increases Sn coordination numbers. Minimization of energy within the ssNEB 
formalism would allow for some rearrangements of atoms and lowering of the energy barriers along these pathways. Indeed, the energy barriers calculated 
from the direct interpolation along our pathways are about 216 and 124 meV/atom for Pbcn $\rightarrow$ P4$_2$/mnm and Pa3 $\rightarrow$ P4$_2$/mnm,
respectively. The subsequent ssNEB calculations lower the energy barriers in both cases, but while for the Pbcn $\rightarrow$ P4$_2$/mnm the symmetry of the 
initial pathway remains the same with the final barrier of 115 meV/atom, the ssNEB result for the Pa3 $\rightarrow$ P4$_2$/mnm relaxes from the initial pathway to the one that has the 
Pbcn structure as the intermediate. The ssNEB energy barrier for the Pa3 $\rightarrow$ Pbcn transition is 32 meV/atom and is in qualitative agreement with 
the direct interpolation one of 97 meV/atom. Given that the Pa3 structure is the next in the sequence of increasing pressure after the Pbcn, the ssNEB result 
from Fig.~\ref{fig5} explains the appearance of the Pbcn phase mixed with the ground-state rutile phase (P4$_2$/mnm) upon releasing pressure.
Namely, the barrier of 32 meV/atom implies Pa3 will transform to Pbcn fairly quickly, while the Pbcn phase will transform to P4$_2$/mnm at a slower rate. 
Interestingly, just from the coordination analysis and the calculated energy profiles one could conceive the Pa3 $\rightarrow$ Pbcn $\rightarrow$ P4$_2$/mnm 
route instead of the direct Pa3 $\rightarrow$ P4$_2$/mnm.\par

Similarly, the ssNEB result for the Pbca $\rightarrow$ P4$_2$/mnm transformation (see Fig.~\ref{fig5}(d)) is clearly consistent with the conclusions that could be 
drawn form the coordination analysis. Namely, from the coordination of atoms this transformation would undoubtedly be considered slow, and the lowering of 
energy from Pbca would likely proceed along the pathways marked by the arrows in Fig.~\ref{fig4}, that is, either through the Pa3 and then Pbcn, or directly to the Pbcn structure. 
As previously discussed both of these transformations can be considered rapid. This is exactly what the ssNEB predicts would happen. The Pbca would, according to 
ssNEB, transform rapidly into the Pbcn (barrier of 44 meV/atom) and would then continue along the Pbcn $\rightarrow$ P4$_2$/mnm path.\par

As the SnO$_2$ results demonstrate, structure mapping and careful coordination analysis can offer qualitative guidance and accelarate the classification of the polymorphic
transformation into rapid and slow. While replacing the ssNEB results with just the energy profiles calculated from the interpolation of our pathways might be tempting,
one needs to remember that the barriers calculated in this way only represent the upper bounds for the true activation energy. The problem occurs if the upper bound for 
the activation energy is relatively large. We argue here that under these circumstances careful coordination analysis can still be useful in providing qualitative assessment 
as discussed for the transitions denoted by dashed lines in Fig.~\ref{fig4}. Of course, the ssNEB in these cases would provide the ultimate answer.

We further tested our mapping algorithm and the qualitative conclusions based on coordination analysis using SiO$_2$ as another case study (see Supplementary 
Materials for details). The resulting classification based on the coordination of atoms shows fast transformations between $\alpha$- and $\beta$-quartz,
$\alpha$- and $\beta$-cristobalite, and $\alpha$- and $\beta$-tridymite. All other transformations between them and also including high-pressure
moganite and stishovite would be classified as slow (coordination does not depend on the tolerance factor). This entirely geometric- and coordination 
of atoms- based classification reproduces qualitatively well the available knowledge of the SiO$_2$ polymorphs and the kinetics of polymorphic transformations
between them.\par
%
\section{Conclusions}

In conclusion, with the main goal of accelerating the assessment of the kinetics of polymorphic transformations
we developed a general algorithm to map crystal structures onto each other and construct transformation pathways between them.
The algorithm itself is based on the physical principles of minimizing the distance atoms need travel between the 
two structures (diffusionless transformations) and minimizing the change in the first shell coordination of atoms
along the pathway (minimal dissociations of chemical bonds). The developed algorithm reliably reproduces known transformation 
pathways and reveals the critical role the change in coordination of 
atoms plays in identifying the correct mapping. 
Moreover, we find that the change in coordination along the pathway can also be used as a proxy for the 
kinetics of polymorphic transformations in partially ionic systems. Namely, the condition of the first-shell coordination of atoms not
decreasing below the end structures, {\it i.e.},  minimal dissociation of chemical bonds along the pathway, can be used as a signature 
of rapid transformations. In covalent systems the geometric coordination of atoms needs to be replaced by the proper bonding analysis.
This allows qualitative, quick and reliable classification of polymorphic transformations into rapid and slow just from the coordination analysis.
For more quantitative assessments integration of our algorithm with ssNEB calculations is shown to provide a robust computational procedure for 
predicting kinetics of polymorphic transformations. Ultimately, the methods presented here, in combination with structure and property predictions, 
offer a route to identifying novel, realizable and long-lived functional materials among metastable polymorphs.\par
%
%
\section{Methods}\label{methods}
%
All calculations were performed by employing relatively standard DFT computations explained in details elsewhere \cite{vladan_PRB:2012}. 
In short, the PBE form of the exchange-correlation functional \cite{perdew_PRL:1996} was used within the projector augmented wave 
(PAW) method \cite{bloechl_PRB:1994} as implemented in the VASP code \cite{kresse_CMS:1996}. In case of elemental carbon, the
VASP code implementation of the optB86 exchange-correlation functional \cite{klimes_JPCC:2010}, which includes contributions arising 
from van der Waalls interactions, was employed. To generate energy profiles along the mapping (pathway) all initial and final structures 
were fully relaxed (volume, cell shape and atomic coordinates), whereas only the static calculations were performed for the intermediate 
structures along the pathway. \par

The solid state nudged elastic band (ssNEB) calculations were performed using the implementation in the Transition State 
Tools for VASP (VTST) code developed by the Henkelman group at UT Austin.\cite{sheppard_JCP:2012} 
Each ssNEB calculation was initialized with an approximately 20 image band determined by taking an interpolation 
of the entire desired polymorphic transformation generated by the mapping algorithm discussed in this paper. 
The initial pathway is relaxed until the forces on all images were less than 0.01 eV/{\AA}. If any intermediates appeared along the ssNEB, 
these were relaxed using a standard geometry relaxation, and the ssNEB path was split, taking the intermediate as a new end point for 
one ssNEB, and a starting point for another. This was done so that each ssNEB calculation would only have a single local maximum. Images 
were either added or removed from the chain so that each intermediate ssNEB had 10-12 images. Once the NEB images were relaxed, the 
climbing image NEB method was used to better identify the true saddle point.\cite{henkelman_JCP_2:2000}

%
\begin{acknowledgments}

This work was supported as part of the Center for the Next Generation of Materials by Design, an Energy Frontier Research Center 
funded by the U.S. Department of Energy, Office of Science, Basic Energy Sciences. The research was performed using computational 
resources sponsored by the Department of Energy's Office of Energy Efficiency and Renewable Energy and located at the National 
Renewable Energy Laboratory.
\end{acknowledgments}

%
%
\end{document}